\documentclass[preprint,3p,times,twocolumn]{elsarticle}

\usepackage{newtxtext,newtxmath}
\usepackage{cleveref}
\usepackage{caption}
\usepackage{array}
\usepackage{makecell}
\usepackage{subcaption}

\journal{Engineering Applications of Artificial Intelligence}

\begin{document}


\begin{frontmatter}

\title{Condition monitoring of wind turbine blades via learning-based methods}

\author[1]{Giovanni Zaniboni}
\ead{giovanni.zaniboni@mail.polimi.it}
\affiliation[1]{organization={Department of Electronics, Information and Bioengineering (DEIB)},
            addressline={Via Ponzio 34/5},
            city={Milano},
            postcode={20133},
            country={Italy}}
\author[2]{Alessio Dallabona\corref{cor1}}
\ead{aldall@dtu.dk}
\cortext[cor1]{Corresponding author.}
\affiliation[2]{organization={Department of Electrical and Photonics Engineering, Technical University of Denmark},
            addressline={Elektrovej 326},
            city={Lyngby},
            postcode={2800},
            country={Denmark}}
\author[3]{Johnny Nielsen}
\ead{jynie@vestas.com}
\affiliation[3]{organization={Vestas Wind System A/S},
            addressline={Hedeager 42},
            city={Aarhus},
            postcode={8200},
            country={Denmark}}
\author[2]{Dimitrios Papageorgiou}
\ead{dimpa@dtu.dk}

\begin{abstract}
This paper addresses the topic of condition monitoring of wind turbine blades and presents a learning-based approach to fault detection. The proposed scheme utilises Principal Components Analysis and Autoencoders to derive data-driven models from root-bending moment and other measurements. The models are trained with real data obtained from a fault-free wind turbine, and then validated on data corresponding to unknown health condition. Online test statistics, employing static thresholds and Generalized Likelihood Ratio tests, are used on residual signals generated by discrepancies between the actual and reconstructed measurements to detect deviations from nominal operation. The efficacy and effectiveness of the proposed methods are demonstrated using real-life data collected from wind turbines experiencing blade faults.
\end{abstract}


\end{frontmatter}


\section{Introduction}
\label{sec:intro}
The Wind Turbine (WT) market has experienced significant growth in recent years and is projected to continue expanding over the next few decades. Governments, such as the European Commission, are committing substantial resources to transition towards renewable energy sources, aiming for complete carbon neutrality by 2050 \cite{WindEU}. 

As the number of wind turbines continues to rise, there is a growing need to identify new suitable locations for new wind farms. The limited availability of optimal sites is reflected in the trend of constructing taller turbines with bigger blades for making the investment economically viable in less optimal areas. The design of bigger blades requires different solutions, with the consequence of dealing with different fault scenarios that have higher economical impact. Even with the current blade designs state, about $14.9\%$ of the faults on WT are blade failures \cite{CHOU201399}, caused by both endogenous and exogenous factors \cite{s22041627,KONG2023390,en}. Furthermore, blades are the structural element of the turbine most affected by damage, due to their high complexity and prolonged exposition to strong dynamical loads \cite{s22041627}. As a consequence, interest arose in the investigation of condition monitoring and fault detection strategies to be able to detect faults preemptively before they become more severe and thus considerably more costly. A comprehensive review of the research methods is presented in \cite{Ref1,Ref2}.

Model-based Fault Detection (FD) strategies are the most traditional schemes used in wind turbines and their subsystems \cite{Ref3}. Starting form a linearised model of the system, residuals are commonly generated starting from Kalman filters and the collected data \cite{Gustafsson2000AdaptiveFA}. In \cite{WEI20083222} and \cite{5571962} a closed-loop scheme that describes the WT model is provided, which was then used for residual generation using Kalman Filters. In the latter, wind speed was taken as a reliable measurement for the design of the Kalman filter. However, it is generally difficult to obtain accurate local wind speed measurements \cite{bianchi_wind_2007}. This issue was addressed in \cite{enevoldsen_condition_2020}, where the authors estimated the wind speed via estimating the aerodynamic torque.

Such limitations of conventional model-based methods can oftentimes be overcome by leveraging data-driven methods that only rely on sensor readings as opposed to estimating unmeasured signals \cite{silvioNN}. Principal Component Analysis (PCA) was implemented in \cite{en9010003} and \cite{8283636} to detect sensors and actuator faults in wind turbines. In \cite{VILLEZ200923}, a fault detection algorithm was developed that monitors the data collected online from a plant, whose model was built using Principal Component Analysis. The authors in \cite{BAKDI2019546} presented a method for sensor and actuator fault detection and proposed a way to address the linearity of the PCA by performing multiple model training stages. During each stage, data were collected from a different Operating Region of the WT. A different data-driven fault detection approach was proposed in \cite{novelmethod}, where an alarm signal was computed as a safety threshold on the root bending moments. 

Neural network have also been explored because of their capability of dealing with underlying nonlinear functions. Autoencoders (AE) were applied in \cite{8059861}, to detect faults occurring on the gearbox and generator. Other works have continued on this path, training the autoencoders to recognise different faults happening on the WT, such as \cite{article}, and \cite{https://doi.org/10.1002/acs.3685}. An AE model trained with Long Short-Term Memory (LSTM) layers was proposed in \cite{9327166} to recognise when a blade cracks.

Although data-driven methods for fault detection in wind turbines has been studied in the recent literature, dedicated studies on condition monitoring of the blades has been sparsely reported. This paper explicitly focuses on the design and application of learning-based fault diagnosis scheme for detecting blade faults in wind turbines. The main contributions of this paper are
\begin{itemize}
\item Derivation of data-driven models for reconstruction of root-bending moments real measurements.
\item Design of a condition-monitoring framework utilising two versions of data-driven models, namely on based on PCA and one on autoencoders.
\item Integration of statistical change detection into the learning-based framework for robust diagnosis of anomalies.
\item Experimental validation of models and comparative assessment of the condition monitoring methods using data from real-life wind turbines.
\end{itemize}

The rest of this paper is structured as follows: Section 2 describes the characteristics of the wind turbine from which the data was collected and presents the fault scenarios that will be studied. A brief introduction to the methods is presented in Section 3. In Section 4 the details about the design of the different models are presented, and the experimental results obtained by applying the developed algorithms are summarised in Section 5. Finally, conclusions are drawn in Section 6.

\section{System description and problem statement}\label{sec:prob_set}

\begin{figure*}[h!]
    \centering
    \includegraphics[width=\textwidth]{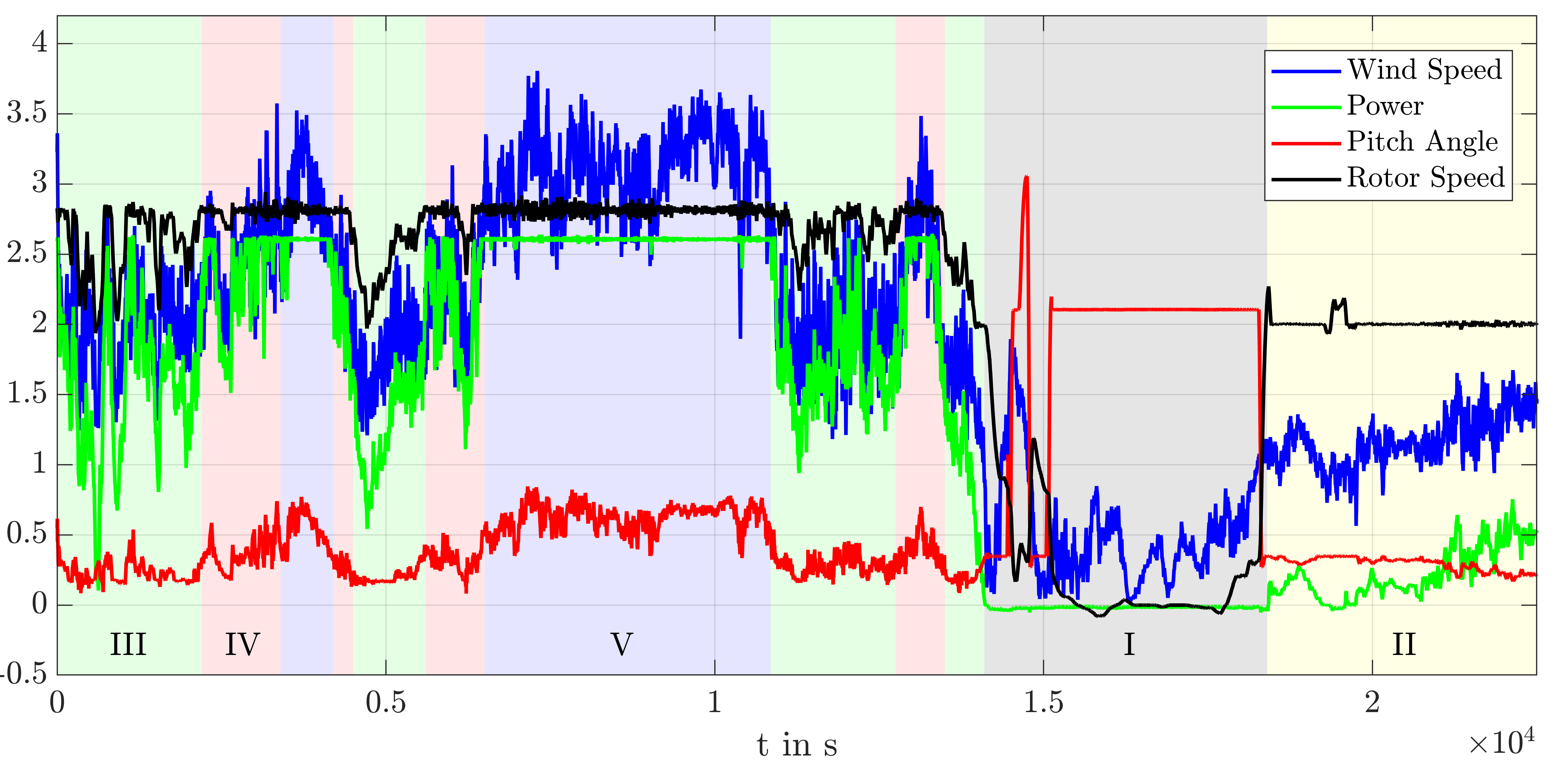}
    \caption{Normalized data for determining the current operating region.}
    \label{fig:op_reg}
\end{figure*}

All the data that were used for the scope of this paper were provided by Vestas Wind System A/S, and were collected from a 2.2 MW variable-speed, variable-pitch horizontal axis wind turbine. The signals that are available from the measurements, and that will be used to train the models are:
 \begin{enumerate}
     \item Flap bending moments acting on each blade (3 measurements)
     \item Edge bending moments acting on each blade (3 measurements)
     \item The Rotor speed
     \item The Wind speed
     \item The Grid power
     \item Pitch angles of each blade (3 measurements).
 \end{enumerate}
Such signals have been chosen because of their relevance, both in terms of being able to display a change under blade fault and for determining the wind turbine operating point correctly. In fact, according to the wind speed, the power output and the rotor speed, 5 different operating regions are defined, where the turbine is operated with different control strategies, and thus is in general characterized by a different underline dynamics. Such signals, together with the consequent blade pitch angle are normalized and presented in Figure~\ref{fig:op_reg}, highlighting the different operating regions at which the turbine is operating at.

A long recording with the turbine not subject to fault will be utilized for fitting the learning-based models. Such data are normalized and split into training, validation and testing for designing and verifying the design under nominal conditions. Some faulty data are available for the same turbine subject to different blade faults. These data were not accounted for in the design stage and are used to test the models under real-life faulty scenarios.

The final goal is to create a condition monitoring framework for detecting blade faults, which starting from the listed measurements is able to detect anomalies by finding discrepancies in their reconstructions.

\subsection{Failure Modes}\label{sec:fail_mode}

The fault scenarios that were considered in the investigation can be divided in three main categories: 

\begin{enumerate}
\item Two common phenomenon that might affect the blade are icing, caused by the formation of ice because of low temperatures, as stated in \cite{s22041627}, and the formation of cracks along the surface of the blades. Such events could entail a rather sudden change in the moment of inertia of the blade or its aerodynamic properties, hence showing a sudden jump in the measurement of the root bending moments, as stated in \cite{lee_transformation_2015}. Hence, they were simulated on the available data from the healthy turbine by adding an offset to the original bending moment measurements. A bias in the measurement could also be caused by a fault occurring in the sensor itself. The fault described by the biased root bending moment measurement is hence representing both a faulty operation derived from normal physical phenomenon and a the corruption of a critical sensor for the turbine.
\item Another common fault is represented by the case of a sensor measurement being stuck because of its incorrect setup on the blade, or by defects in the sensor itself. This kind of fault was simulated by changing the measured bending moment signal to become constant after the fault occured.
\item Finally, it was decided to use real-life data collected from a blade undergoing a failure. For this reason, Vestas Wind Systems A/S provided some data collected with the turbine experiencing a fault of unknown origin, representing a real-life use case of a condition monitoring algorithm.
\end{enumerate}

\section{Methods and theoretical overview}

Both the implemented Fault Detection methods rely on the measurements provided by the different wind turbine sensors, to develop a data-driven model of the healthy system by finding the main features between the collected data. This first stage is the training stage and is performed offline. 

Once the model is trained, the testing Stage is performed online: new data are collected from the wind turbine, into an unknown state, and then reconstructed by the trained model. The fault detection algorithm is obtained by computing a suitable statistic to compare the data reconstructed by the trained model and the real data measured by the sensors (usually by means of Square Prediction Error or Mean Average Error), and by comparing the result with the fault-free case obtained in the training stage by means of a Generalized likelihood ratio (GLR) test. If the two statistics are similar, then the GLR test assumes that the new data are from a healthy plant, while if they are showing strong differences, it detects a fault. Two approaches are proposed (PCA and Autoencoders), differing both in the model structure and in the training method.

\subsection{Introduction to PCA}\label{sec:pca_intro} 

PCA is an unsupervised, linear method for dimensional reduction in large multivariate datasets. It transforms a normalized data matrix $\pmb{X} \in \mathbb{R}^{n \times m}$ (with $n$ samples and $m$ features) into a lower-dimensional space by applying singular value decomposition (SVD) to its covariance matrix:
\begin{equation}
\pmb{S} = \frac{1}{n-1}\pmb{X}^T\pmb{X} = \pmb{P\Lambda P}^T.
\end{equation}
Here, $\pmb{P}$ contains the eigenvectors, and $\pmb{\Lambda}$ is a diagonal matrix with eigenvalues in descending order. Traditional PCA, however, is limited by its assumption of linear and time-invariant relationships between features. To address nonlinearities and capture temporal dependencies in datasets, Dynamic PCA (dPCA) was introduced. dPCA extends the conventional PCA framework by incorporating a time window $\omega$, resulting in an expanded data matrix $\pmb{X_d} \in \mathbb{R}^{n-(\omega-1) \times \omega m}$:
\begin{equation}
\pmb{X_d}(k) = \begin{bmatrix}
      \pmb{X}(k-\omega+1) & \ldots & \pmb{X}(k) 
      \end{bmatrix},
\end{equation}
where $k$ ranges from $1$ to $n-(\omega-1)$.

Upon applying PCA to $\pmb{X_d}$, the significant principal components are retained based on their eigenvalues. Each row $\pmb{X_d}(k)$ can be decomposed into its principal subspace projection $\pmb{\hat{x}}(k)$ and the residual $\pmb{e}(k)$:
\begin{equation}
\pmb{X_d}(k) = \pmb{\hat{x}}(k) + \pmb{e}(k).
\end{equation}

The model's effectiveness is quantified using the Squared Prediction Error (SPE) statistic for each data point, providing a measure of deviation from the modelled behaviour. The SPE statistic measures, for every row of $\pmb{X_d}$,
\begin{equation}
\begin{aligned}
SPE(k) &= \sum_{i=1}^{m\omega}\left(\hat{X}_d^{(i)}(k)-X_d^{(i)}(k)\right)^2\\
&=\pmb{X_d}(k)(\pmb{I}-\pmb{\hat{P}\hat{P}}^T)(\pmb{I}-\pmb{\hat{P}\hat{P}}^T)^T\pmb{X_d}(k)^T
\end{aligned}
\end{equation}
which is the sum of all the squared components of $\pmb{e}(k)$. This dynamic approach is particularly beneficial for systems like wind turbines, where operational dynamics introduce significant nonlinearity and time-dependent behavior in sensor data.

\subsection{Introduction to Autoencoders}

Autoencoders (AEs) are a type of neural network designed for unsupervised learning, primarily aimed at data reconstruction. They consist of two main components: an encoder and a decoder. The encoder compresses the input data $\pmb{X}(k) \in \mathbb{R}^m$ into a lower-dimensional latent space $\pmb{Z}(k) \in \mathbb{R}^q$, through a series of transformations:
\begin{equation}
\pmb{E}_i(k) = \begin{cases}
\varphi_1^e(\pmb{W}_1^e\pmb{X}(k)+\pmb{b}_1^e), & \text{if } i=1 \\
\varphi_i^e(\pmb{W}_i^e\pmb{E}_{i-1}(k)+\pmb{b}_i^e), & \text{otherwise}
\end{cases}
\end{equation}
and
\begin{equation}
\pmb{Z}(k) = \varphi^z(\pmb{W}^z\pmb{E}_\zeta(k)+\pmb{b}^z),
\end{equation}
where $\zeta$ denotes the number of encoder layers. The decoder, in a similar fashion, is then leveraged to reconstruct the original input from these latent variables, culminating in the output $\pmb{\hat{X}}(k)$.

The performance of an AE is commonly evaluated by the mean squared error (MSE) between the original and reconstructed data, minimized during training:
\begin{equation}
MSE(\pmb{X}, \pmb{\hat{X}}) = \frac{1}{n} \sum_{k=1}^{n} \|\pmb{X}(k) - \pmb{\hat{X}}(k)\|^2.
\end{equation}
Given the non-linear and time-variant nature of wind turbine (WT) data, standard AEs are often insufficient. Therefore, the model is extended with recurrent structures such as Long Short-Term Memory (LSTM) layers to capture temporal dependencies, and convolutional layers for spatial feature extraction in data. Additionally, a dynamic expansion of the input data matrix $\pmb{X}$ to $\pmb{X_d} \in \mathbb{R}^{(n-(\omega-1)) \times (m\omega)}$ allows the AE to adapt to varying operational conditions by expanding the input space temporally. These adaptations are essential for modeling the complex dynamics of WT systems under different environmental conditions.

\subsection{General Likelihood Ratio Test}\label{GLR} 

The General Likelihood Ratio Test (GLRT) is employed to determine whether a signal adheres to a specified probabilistic model under null hypothesis $\mathcal{H}_0$, or if it instead follows a different distribution under alternative hypothesis $\mathcal{H}_1$ characterized by a different mean value. Specifically, for signals that normally distribute, the GLRT is computed as:
\begin{equation}\label{eq:glr}
g(k) = \frac{1}{2\sigma^2M} \max_{k-M+1 \leq j \leq k} \left(\sum_{i=k-M+1}^k (z(i) - \mu_0)^2\right).
\end{equation}
where $z(i)$ represents the realization of the signal, and $\mu_0$ and $\sigma$ are the mean and standard deviation of the signal under $\mathcal{H}_0$. The calculation within a running window $M$ is leveraged to enhance computational efficiency. The statistic $g(k)$ is compared against a threshold $h$ to decide between $\mathcal{H}_0$ (non-faulty behavior) and $\mathcal{H}_1$ (fault indication). $M$ and $h$ are design parameters that can be calculated starting from the statistical property of the signal under $\mathcal{H}_0$ and the expected variation under $\mathcal{H}_1$ \cite{book}.

\section{Design of condition monitoring framework}

\begin{figure*}[h!]
    \centering
    \includegraphics[width=0.8\textwidth]{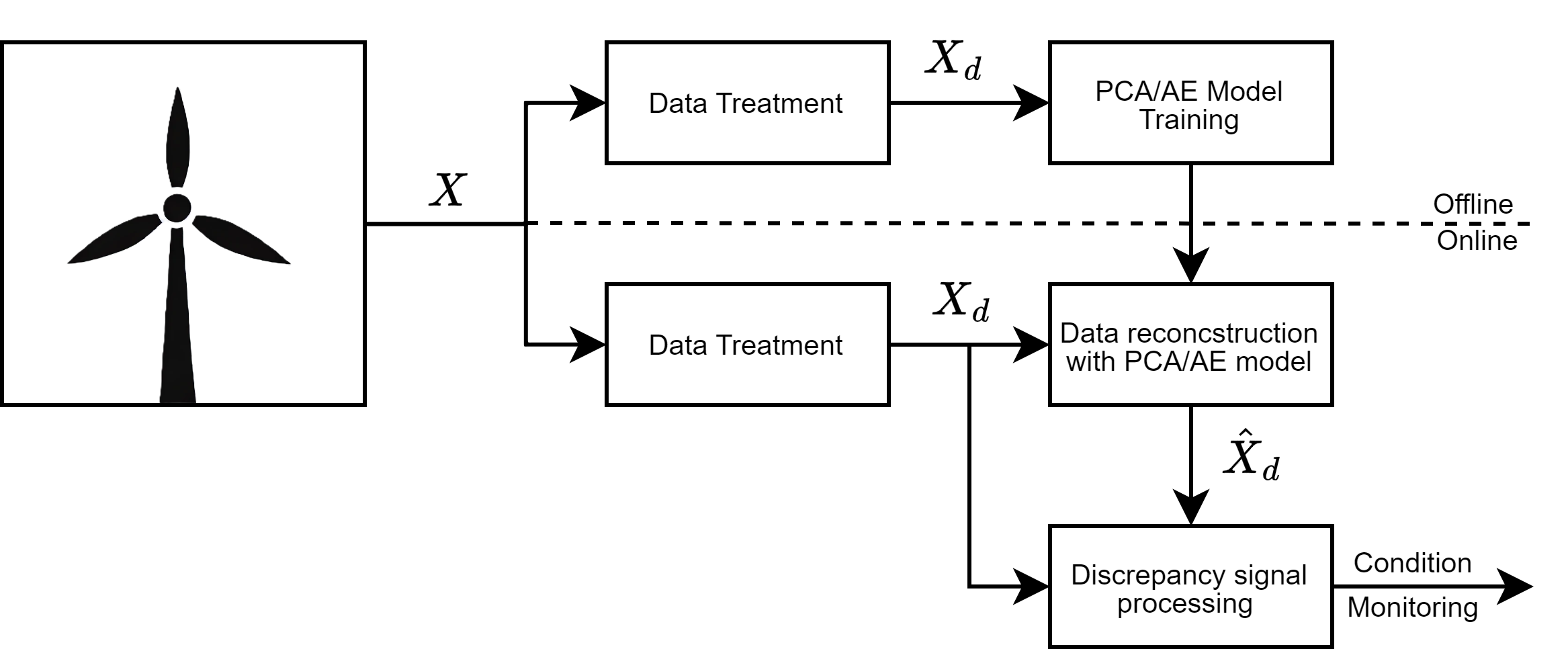}
    \caption{Block diagram for the condition monitoring framework based on the wind turbine measurements and the reconstructed data.}
    \label{fig:general_diagram}
\end{figure*}

The condition monitoring framework consist of an offline training stage and an online implementation one. The former involves a pre-processing of the data presented in Section~\ref{sec:prob_set} for training the models, models that are then utilized in the former for reconstructing new real-time data that are pre-processed in the same fashion. The comparison between the original and reconstructed data is processed for assessing the presence of faults. A summary of the entire framework is presented in Figure~\ref{fig:general_diagram}.

\subsection{PCA} \label{PCA: Design}

Because of the limitation of standard PCA introduced in Section~\ref{sec:pca_intro}, a dPCA model is developed. All the signals defined in Section~\ref{sec:prob_set} were used for training the model. Thus, every datapoint in the training dataset $\pmb{X}(k)$ measured at time instant $k$ had $12$ components. The collected data matrix for training the model is $\pmb{X_0} \in \mathbb{R}^{n\times12}$, where $n$ would change according to the size of the training dataset chosen for each operating region. To address PCA's constraint of being able to only identify linear correlations, the data are split for the different operating regions and equal number of model is trained. The first operating region is not considered, since it is not of interest developing a model for detecting faults in a non-operational turbine. The process of identifying distinct training datasets involved analyzing all recorded data and comparing the behaviour of various variables (e.g. the rotor speed, grid power and pitch angles) to properly separate all the operating regions. 

As a final parameter needed to build the training datasets for dPCA, the time window was set as $\omega = 100$, which corresponds to a time of $10$ seconds. The value has been chosen for being able to show enough variation in the signals, for allowing the model to find time varying correlations between the data and approximating nonlinearities in the relationships between the various measured variables.

After performing SVD on the normalized data matrix, the value $l$ for dimensionality reduction was chosen according to the Cumulative Variance (CV), defined as
\begin{equation}
CV = \frac{\sum_{i=1}^{l} \lambda_i}{\sum_{i=1}^{m\omega}\lambda_i},
\end{equation}
where $\lambda_i, 1\leq i \leq m\omega$ is the i-th orthonormal loading found by PCA. The resulting numbers of principal components for a chosen CV of $ 90\% $ applied to all four operating regions are summarised in Table~\ref{tab: Principal Components}.

\begin{table}[h]
\centering
\begin{tabular}{|c|c|c|c|c|}
\textbf{Zone} & II & III & IV & V \\
\hline
\textbf{$l$} & 4 & 6 & 6 & 8 \\
\end{tabular}
\caption{Table showing the chosen amount of PCs for each of the operating regions}
\label{tab: Principal Components}
\end{table}

\begin{figure}[h!]
    \centering
    \includegraphics[width=1\columnwidth]{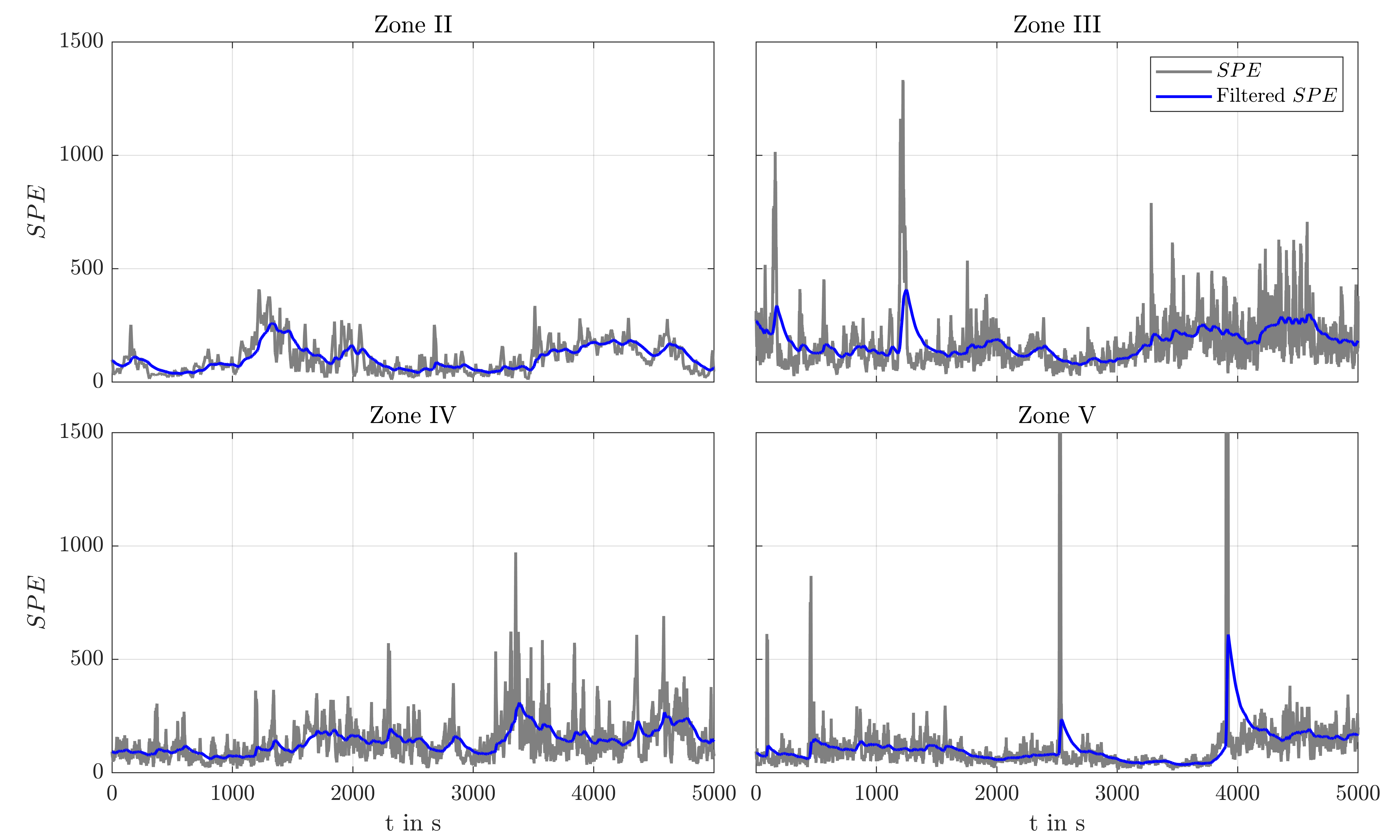}
    \caption{SPE computed over the training datasets of the $4$ operating regions}
    \label{fig: training SPE}
\end{figure}

The resulting Squared Prediction Errors (SPEs) for the reduced models are presented in Figure~\ref{fig: training SPE}.
 
The value is fairly constant except for the presence of spikes, which are probably consequent to the presence of data outliers and the difficulty in having a clear distinction among the several operating regions. To attenuate the presence of the spikes, that could lead to the detection of false alarms, a low-pass filter has been designed. The filtered signal can be used for deriving a threshold with which it will be compared during the online implementation for detecting the presence of a fault. The threshold was computed by setting a probability of a false alarm, in this case $P_F = 1 \%$, and then choosing the threshold as the value such that it is smaller than the $P_F$ percentage of all the computed SPE points of the training dataset.  

\subsection{Autoencoders} \label{Design: AE}
Autoencoders were implemented as an improvement of the PCA approach, with their ability to model the nonlinearities present in the wind turbine dynamics. The AE approach allows, starting from just one training dataset, to obtain a model of the healthy wind turbine that covers all five operating regions, provided that the data chosen for the training stage covers all of them.

In the data collection, for each data point, $9$ features were considered: 
\begin{enumerate}
\item Flap bending moments on all three blades (3 measurements)
\item Edge bending moments on all three blades (3 measurements)
\item Rotor speed
\item Wind speed
\item Grid power.
\end{enumerate}

\begin{figure*}[h!]
\centering
\includegraphics[width=0.8\textwidth]{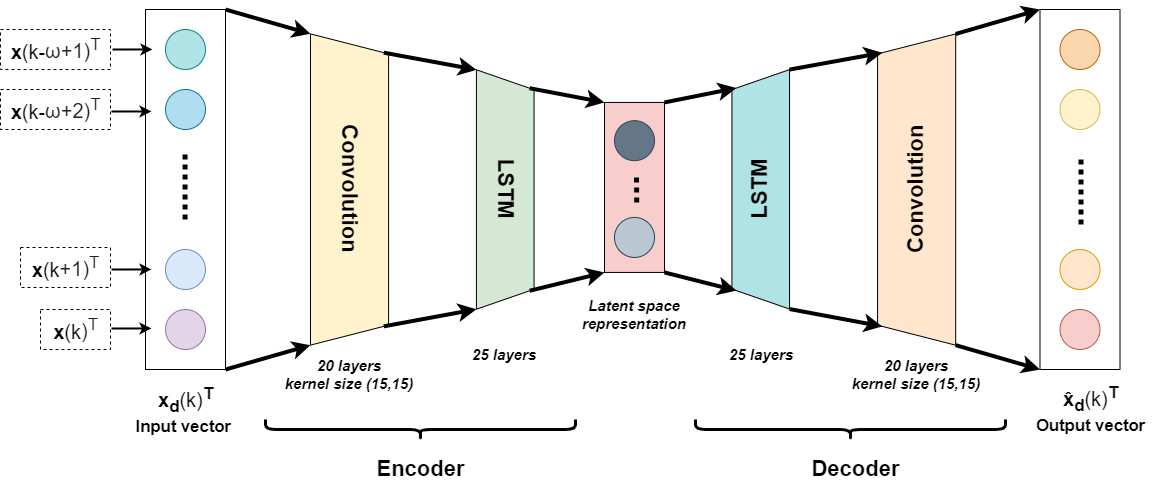} 
\caption{Outline of the autoencoder architecture with both convolutional and LSTM layers. The input vector $\pmb{x_d}(k)$ is collecting the measured data over the sliding window $\omega$, and the AE reconstructs it as $\pmb{\hat{x}_d}(k).$}
\label{fig:AE architecture}
\end{figure*}

The number is reduced in comparison with the PCA model to decrease the computational time, which in the case of 12 features would have been considerably longer.
The time window for including dynamical correlation was set to $10$ seconds, to match the PCA model. In fact, a shorter time window wouldn't allow the coverage of enough variation in the data. Convolutional and LSTM layers have been combined for improving the modeling capabilities, the activation function was chosen to be the hyperbolic tangent. The architecture of the chosen AE, together with the main parameters of interest is outlined in Figure~\ref{fig:AE architecture}.

\begin{figure}[h!]
    \centering
    \includegraphics[width=\columnwidth]{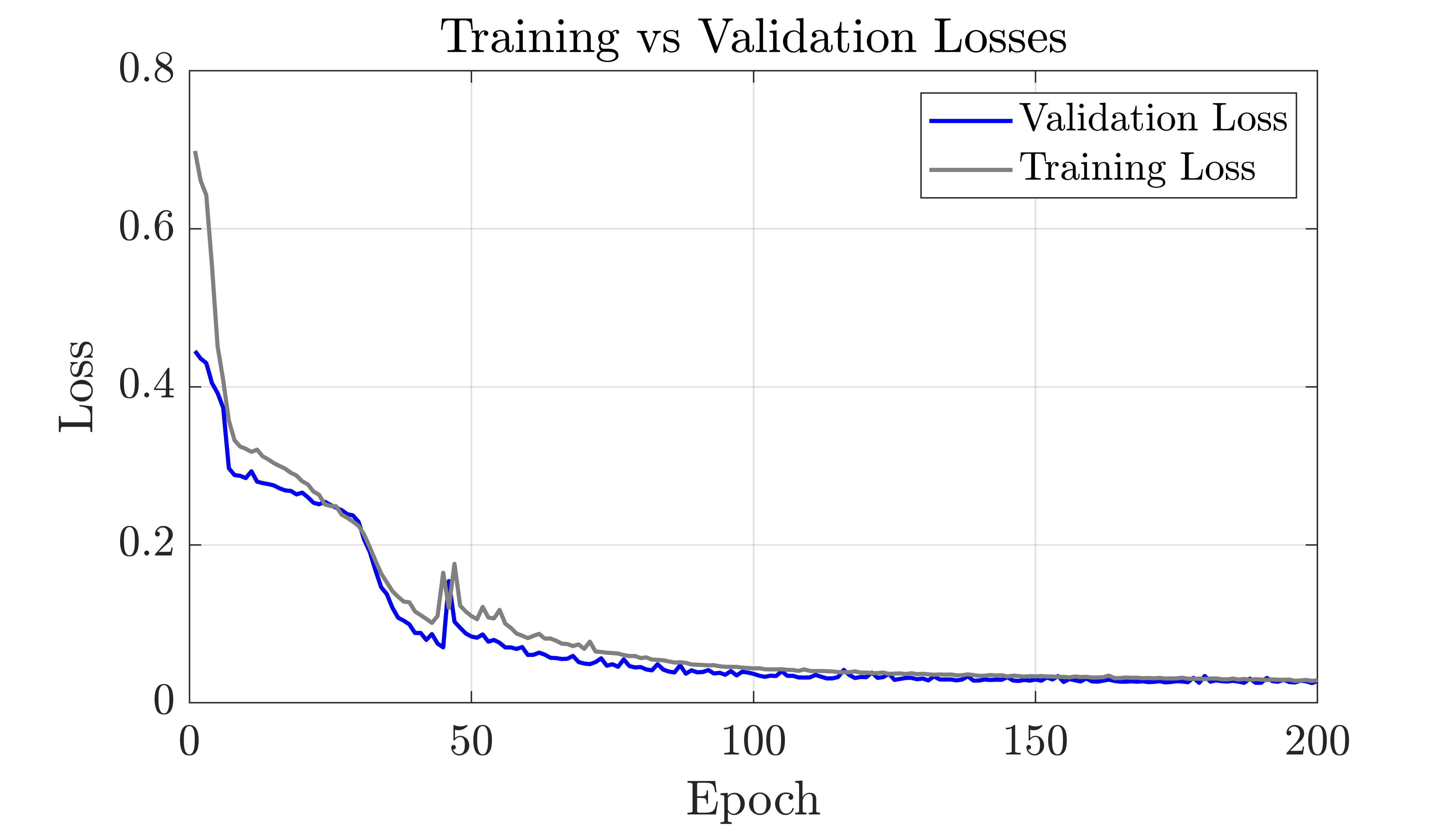}
    \caption{AE Model training and validation}
    \label{fig:epochs}
\end{figure}

The model has been trained with the described dataset. 200 epochs were selected to avoid overfitting, while still being able to model the non-linear relations among the measurements. The results from the training process are stated in Figure~\ref{fig:epochs}.

\begin{figure}[h!]
    \centering
    \includegraphics[width=\columnwidth]{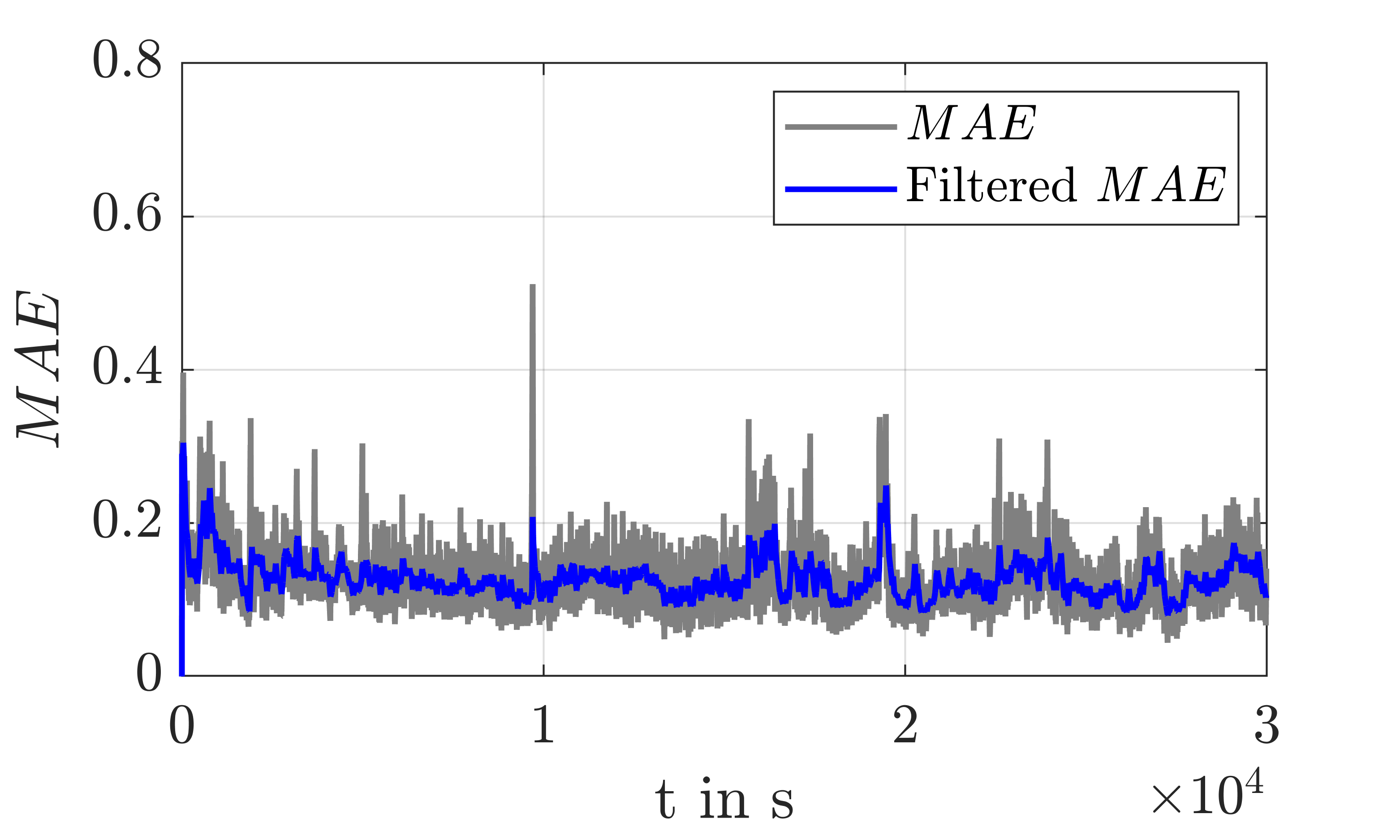}
    \caption{$MAE_{train}$ computed over the training dataset using the architecture with both convolutional and LSTM layers}
    \label{fig:mae train}
\end{figure}

The performance of the trained model was evaluated by calculating the MAE, stated in Figure~\ref{fig:mae train}. Such metrics allowed to choose the current architecture among others, as the best performing one in reconstructing the nominal data. Similarly to the PCA case, the presence of spikes led to the design of a LPF for reducing the probability of false detection, in case a static threshold is applied directly to the current metrics. The threshold can similarly be calculated as it was done previously for the PCA case.

\subsection{GLR Test}

A GLR test is designed for leveraging the statistical properties of the reconstruction error of each model in order to better assess the presence of a change. The natural filtering action consequent from evaluating the signal over a time window and from knowing the standard deviation of the underlying gaussian distribution, replaces the LPF previously designed and improves the signal separation between the non-faulty and faulty case for better assessing the presence of faults.

The test is implemented according to Equation~\ref{eq:glr}, with the $h$ and $M$ parameters being calculated to guarantee a certain probability of detection $P_D = 0.99$ and false detection $P_F = 0.01$ for $\mu_1 = 2\mu_0$.

\section{Experimental results}

Both the PCA and the AE models are tested for the faulty scenarios presented in Section~\ref{sec:fail_mode}. The definition of the experiments for the third category is straightforward, thanks to the availability of faulty data. Two recordings are available, namely Real-life scenario 1 and Real-life scenario 2, with different evolution characteristics. On the other hand, the first two categories require the corruption of (unused) healthy data to mimic the faulty scenarios described in Section~\ref{sec:fail_mode}. Each of the first two scenarios (bias and stuck) are applied to both the edge and the flap bending moment. An overview of the experiment setting for each faulty scenario is presented in Table~\ref{faults}, where $M_{og}(t)$ represents the blade root bending moment measured on the original healthy turbine at time $t$, and $t_f$ the time instant where the fault occurs.

\begin{table}[h]
\centering
\caption{Overview of the faulty data for the different fault scenarios.}
\label{faults}
\begin{tabular}{>{\raggedright\arraybackslash}p{0.42\columnwidth}|>{\raggedright\arraybackslash}p{0.45\columnwidth}}
\hline
\textbf{Fault scenario} & \textbf{Description} \\ \hline
Flap bending moment offset $\Delta_M$ & $M = M_{og} +\Delta_M, $ $\Delta_M = 1 \times 10^6 \, \text{Nm}$ \\ \hline
Edge bending moment offset $\Delta_M$ & $M = M_{og} +\Delta_M, $ $\Delta_M = 1 \times 10^6 \, \text{Nm}$ \\ \hline
Flap bending moment stuck to $M_{st}$ & $M=M_{st} = 0.8 M_{og}(t_f)$ \\ \hline
Edge bending moment stuck to $M_{st}$ & $M=M_{st} = 0.8 M_{og}(t_f)$ \\ \hline
Real-life scenario 1: catastrophic fault & Sudden jump in the flap \newline bending moment signal \\ \hline
Real-life scenario 2: slower-forming & Exponential increase in \newline the flap bending moment \\ \hline
\end{tabular}
\end{table}

\begin{figure*}[h!]
    \centering
    \includegraphics[width=1\textwidth]{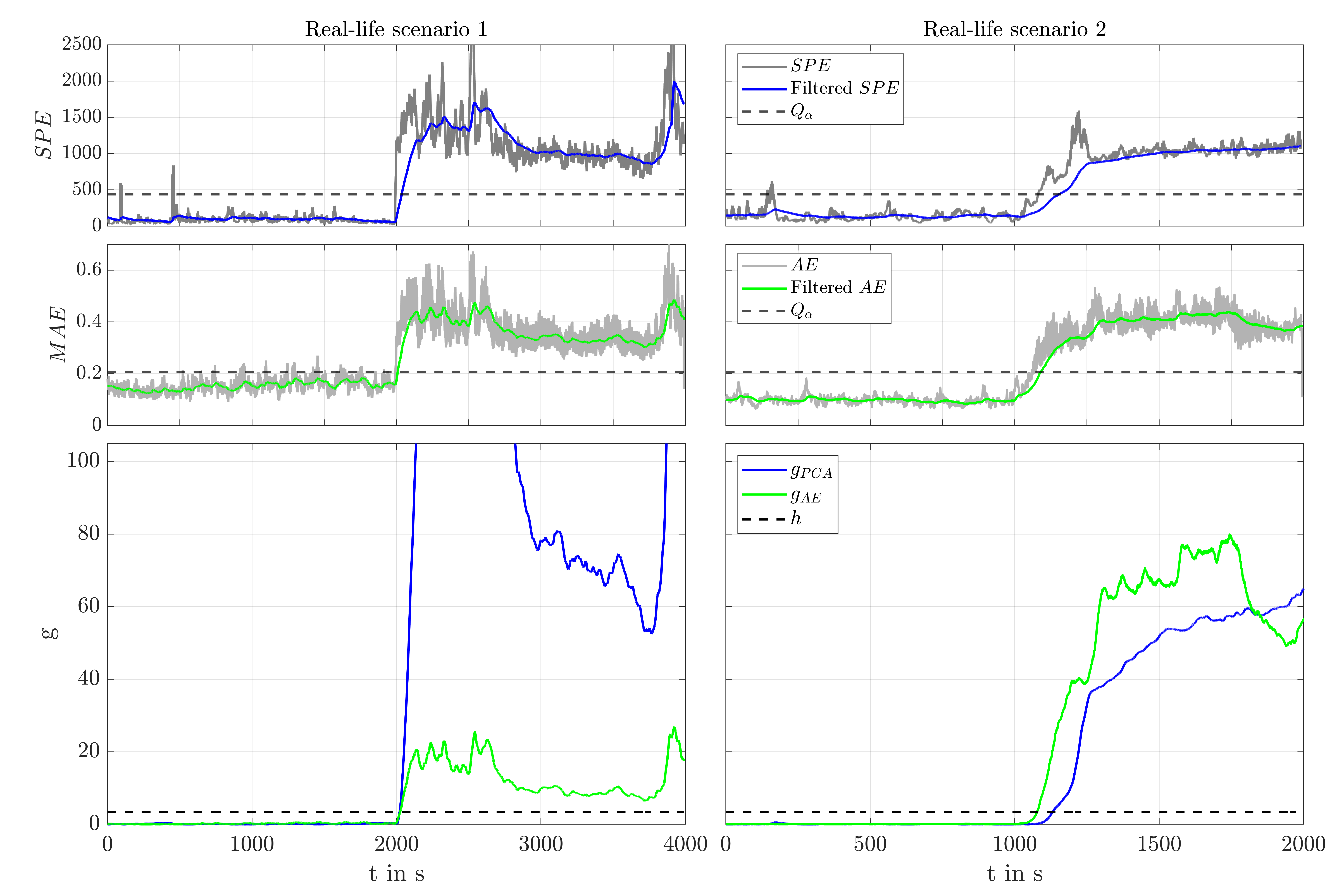}
    \caption{Results from applying algorithms on real faulty data. Comparison between reconstruction error and application of the GLR}
    \label{fig:res_real}
\end{figure*}

Starting from the real-life scenarios, the reconstruction errors and the results of applying the GLR are presented in Figure~\ref{fig:res_real}. 
For the two scenarios under analysis, the faults are detected by both the PCA and AE models. This is observed when applying a threshold to the filtered reconstruction errors (SPE and MAE for the two methods, respectively) as well as when using the GLR, where the signal separation becomes more distinct.

The drawback of using real faulty data, is that the experiment is not repeatable. As a consequence, the statistical confidence of the methods can't be evaluated.

\begin{figure*}[h!]
    \centering
    \includegraphics[width=1\textwidth]{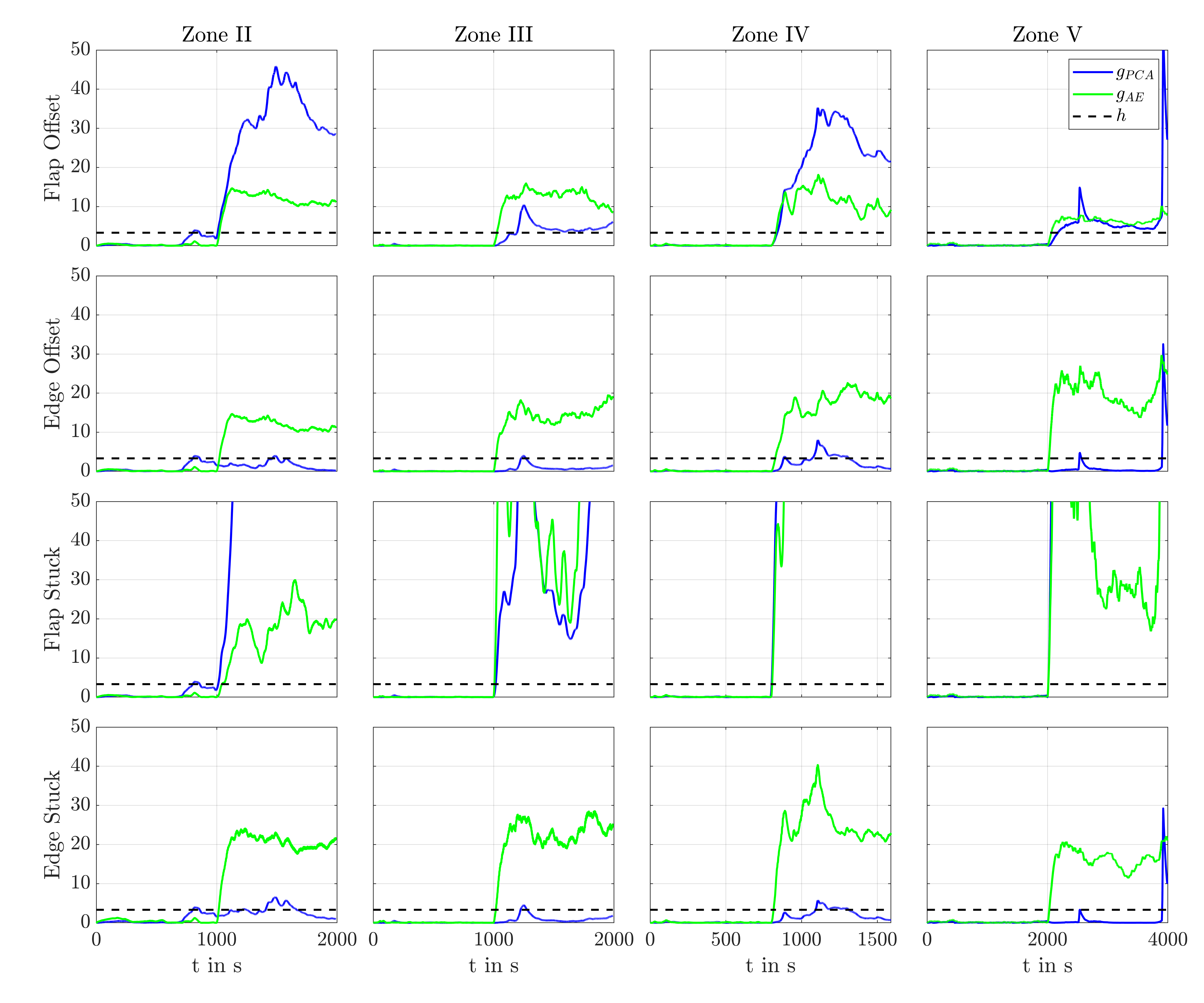}
    \caption{Results from applying algorithms on faulty data obtained by injecting sensor faults. Comparison of change detection properties between PCA and AE.}
    \label{fig:res_sens}
\end{figure*}

The same models are applied to the (new) healthy data corrupted with sensor faults, representing Scenarios 1 and 2. The faults are injected in each operating region, since the PCA requires to use a different model for each of them. The same testing scheme is applied to the AE, to test its capability of capturing the global nonlinear model. Because of the improved detection performance obtained with the GLR (thanks to its signal separation capability), the direct threshold methods results are omitted in the qualitative results presented in Figure~\ref{fig:res_sens}. The capability of the AE of learning the nonlinear features of the model is reflected in a better performance over all the possible faults. In particular, the PCA is in general not able to clearly detect the fault associated with the Edge-bending-moment sensor, while it is still able to detect faults in the flap-bending moment sensor by leveraging the correct model according to the operating region the wind turbine is working at. Even under such conditions, the AE is better performing by ensuring strong detectability and no false alarms in the tests that have been performed.

\subsection{Detection statistical properties}

In order to evaluate the performance of the proposed methods from a quantitative point of view, the possibility of manually corrupting healthy data in the first two faulty scenarios allows to have repeated experiments for each of them, and hence to derive some statistical properties for the defined architectures.

The metrics which is selected is the classic one from fault diagnosis, involving the probability of false alarms and the probability of detection \cite{book}. The first refers to the amount of cases in which the threshold is surpassed before the fault is occurring, the latter to the number of crossing under fault, further distinguishing the two cases of strong and weak detection, where the signals is permanently surpassing the threshold or not, respectively. The total number of detection (without distinguishing between strong and weak) is calculated as well.

The same fault is injected under the same conditions for around 100 times, in order to obtain a precision in the probability percentages close to the unit. In the case of the PCA, where different operating regions are present, the total number of simualtions is set to be around 100 as well, with the number of simulations per operating regions selected according to data availability. Every single simulation lasts 900s, and the fault is injected at 300s.

\begin{figure*}[h!]
    \centering
    \includegraphics[width=1\textwidth]{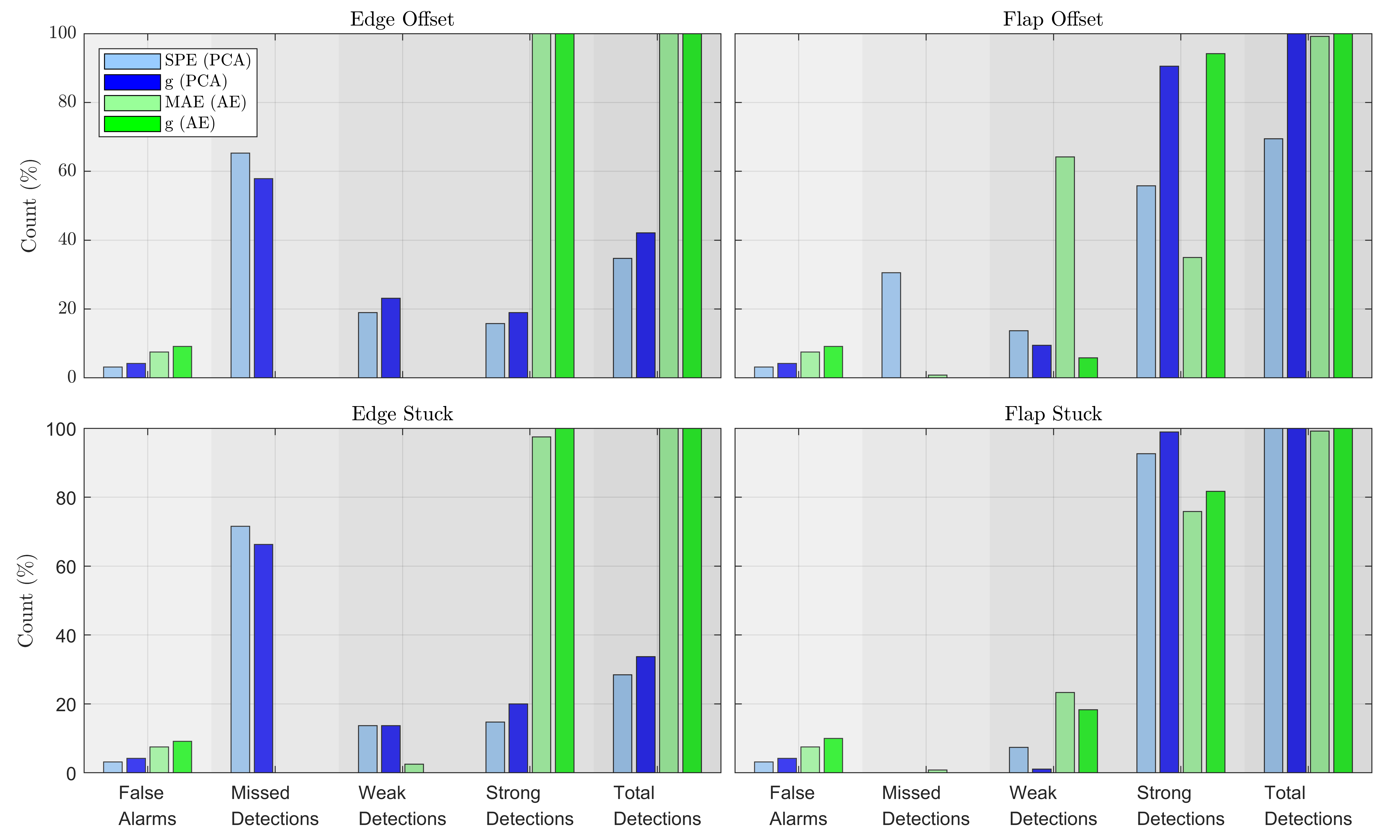}
    \caption{Statistical properties of detection for PCA and AE models.}
    \label{fig:res_stat}
\end{figure*}

The results are presented in Figure~\ref{fig:res_stat}, for both PCA and AE and for both the direct threshold method and the GLR test.

As it was noted from Figure~\ref{fig:res_sens}, both models allow to always detect faults in the flap bending moment sensor. Missed detections are only present with the direct threshold for both of them. The case of the edge bending moment also confirms the outcome pf the previous analysis, with the PCA model showing a high percentage of missed detections with both detection methods (SPE and GLR).

The false alarm rate is generally higher for the AEs and when applying the GLR to the solution. The threshold may be varied to lower the value, at the cost of risking deterioration of the detection performance.

While the AE generally perform better at detection than the PCA, this is not verified for the case where the stuck flap bending moment, where the AE experiences a higher number of weak detections. However, this may be associated to the transient phase, after which the change is strongly detected.

\section{Conclusions}

The lack of investigations focusing on detecting blade faults in wind turbine lead to the definition of a condition monitoring framework based on learning based-models, trained to reconstruct a set of measurements under healthy conditions and showing discrepancies otherwise. Such architecture allows to detect abnormal behaviour in the blades for condition monitoring purposes.

Starting with pre-processed real-life data from a healthy wind turbine, both a linear model based on PCA and a nonlinear model based on autoencoders were trained to reconstruct the input data. The trained models were then tested on faulty data. Although both PCA and autoencoder-based models demonstrated their capability in detecting faults in wind turbine blades, the autoencoder-based approach proved to be the most effective thanks to its ability of dealing with all the tested fault scenarios and by leveraging a single model for all the operating regions. The statistical properties over a higher number of experiments confirmed the qualitative results and provided more insight on other properties, such as false detections, weak/strong detections, etc. Both models have a detection rate close to 100$\%$ for faults related to the flap bending moment, while satisfactory detection performance for edge bending moment faults are only achieved with the AE model.

Future work involves optimizing the algorithm for real-time implementation and testing different model architectures to improve detection performance, and being able to isolate the root causes.

\bibliographystyle{elsarticle-num} 
\bibliography{cas-refs}

\begin{thebibliography}{10}
\expandafter\ifx\csname url\endcsname\relax
  \def\url#1{\texttt{#1}}\fi
\expandafter\ifx\csname urlprefix\endcsname\relax\def\urlprefix{URL }\fi
\expandafter\ifx\csname href\endcsname\relax
  \def\href#1#2{#2} \def\path#1{#1}\fi

\bibitem{WindEU}
E.~Commission, D.-G. for Climate~Action, Going climate-neutral by 2050 – A
  strategic long-term vision for a prosperous, modern, competitive and
  climate-neutral EU economy, Publications Office, 2019.
\newblock \href {https://doi.org/doi/10.2834/02074}
  {\path{doi:doi/10.2834/02074}}.

\bibitem{CHOU201399}
J.-S. Chou, C.-K. Chiu, I.-K. Huang, K.-N. Chi, Failure analysis of wind
  turbine blade under critical wind loads, Engineering Failure Analysis 27
  (2013) 99--118.

\bibitem{s22041627}
M.~Civera, C.~Surace,
  \href{https://www.mdpi.com/1424-8220/22/4/1627}{Non-destructive techniques
  for the condition and structural health monitoring of wind turbines: A
  literature review of the last 20 years}, Sensors 22~(4) (2022).
\newblock \href {https://doi.org/10.3390/s22041627}
  {\path{doi:10.3390/s22041627}}.
\newline\urlprefix\url{https://www.mdpi.com/1424-8220/22/4/1627}

\bibitem{KONG2023390}
K.~Kong, K.~Dyer, C.~Payne, I.~Hamerton, P.~M. Weaver, Progress and trends in
  damage detection methods, maintenance, and data-driven monitoring of wind
  turbine blades – a review, Renewable Energy Focus 44 (2023) 390--412.
\newblock \href {https://doi.org/https://doi.org/10.1016/j.ref.2022.08.005}
  {\path{doi:https://doi.org/10.1016/j.ref.2022.08.005}}.

\bibitem{en}
W.~Wang, Y.~Xue, C.~He, Y.~Zhao,
  \href{https://www.mdpi.com/1996-1073/15/15/5672}{Review of the typical damage
  and damage-detection methods of large wind turbine blades}, Energies 15~(15)
  (2022).
\newblock \href {https://doi.org/10.3390/en15155672}
  {\path{doi:10.3390/en15155672}}.
\newline\urlprefix\url{https://www.mdpi.com/1996-1073/15/15/5672}

\bibitem{Ref1}
H.~Badihi, Y.~Zhang, B.~Jiang, P.~Pillay, S.~Rakheja, A comprehensive review on
  signal-based and model-based condition monitoring of wind turbines: Fault
  diagnosis and lifetime prognosis, Proceedings of the IEEE 110~(6) (2022)
  754--806.
\newblock \href {https://doi.org/10.1109/JPROC.2022.3171691}
  {\path{doi:10.1109/JPROC.2022.3171691}}.

\bibitem{Ref2}
A.~Fekih, H.~Habibi, S.~Simani,
  \href{https://www.mdpi.com/1996-1073/15/19/7186}{Fault diagnosis and fault
  tolerant control of wind turbines: An overview}, Energies 15~(19) (2022).
\newblock \href {https://doi.org/10.3390/en15197186}
  {\path{doi:10.3390/en15197186}}.
\newline\urlprefix\url{https://www.mdpi.com/1996-1073/15/19/7186}

\bibitem{Ref3}
A.~Dallabona, M.~Blanke, H.~Pedersen, D.~Papageorgiou, Fault diagnosis and
  prognosis capabilities for wind turbine hydraulic pitch systems, Mechanical
  Systems and Signal Processing 224 (Jan. 2025).
\newblock \href {https://doi.org/10.1016/j.ymssp.2024.111941}
  {\path{doi:10.1016/j.ymssp.2024.111941}}.

\bibitem{Gustafsson2000AdaptiveFA}
F.~K. Gustafsson, Adaptive filtering and change detection, 2000.

\bibitem{WEI20083222}
X.~Wei, M.~Verhaegen, T.~{van den Engelen}, Sensor fault diagnosis of wind
  turbines for fault tolerant, IFAC Proceedings Volumes 41~(2) (2008)
  3222--3227, 17th IFAC World Congress.

\bibitem{5571962}
X.~Wei, L.~Liu, Fault estimation of large scale wind turbine systems, in:
  Proceedings of the 29th Chinese Control Conference, 2010, pp. 4869--4874.

\bibitem{bianchi_wind_2007}
F.~Bianchi, H.~De~Battista, R.~J. Mantz, Wind Turbine Control Systems:
  Principles, Modelling and Gain Scheduling Design, 2007.

\bibitem{enevoldsen_condition_2020}
T.~T. Enevoldsen, R.~Galeazzi, D.~Papageorgiou, C.~Jeppesen, Condition
  monitoring for single-rotor wind turbine load sensors in the full-load
  region, IFAC-PapersOnLine 53~(2) (2020) 13644--13649, 21st IFAC World
  Congress.

\bibitem{silvioNN}
S.~Farsoni, S.~Simani, P.~Castaldi, Fuzzy and neural network approaches to wind
  turbine fault diagnosis, APPLIED SCIENCES-BASEL 11~(11) (JUN 2021).
\newblock \href {https://doi.org/10.3390/app11115035}
  {\path{doi:10.3390/app11115035}}.

\bibitem{en9010003}
F.~Pozo, Y.~Vidal, Wind turbine fault detection through principal component
  analysis and statistical hypothesis testing, Energies 9~(1) (2016).

\bibitem{8283636}
Y.~Wang, X.~Ma, P.~Qian, Wind turbine fault detection and identification
  through pca-based optimal variable selection, IEEE Transactions on
  Sustainable Energy 9~(4) (2018) 1627--1635.

\bibitem{VILLEZ200923}
K.~Villez, K.~Steppe, D.~J. {De Pauw}, Use of unfold pca for on-line plant
  stress monitoring and sensor failure detection, Biosystems Engineering
  103~(1) (2009) 23--34.

\bibitem{BAKDI2019546}
A.~Bakdi, A.~Kouadri, S.~Mekhilef, A data-driven algorithm for online detection
  of component and system faults in modern wind turbines at different operating
  zones, Renewable and Sustainable Energy Reviews 103 (2019) 546--555.

\bibitem{novelmethod}
K.-Y. Oh, J.-Y. Park, J.-S. Lee, B.~I. Epureanu, J.-K. Lee, A novel method and
  its field tests for monitoring and diagnosing blade health for wind turbines,
  IEEE Transactions on Instrumentation and Measurement 64~(6) (2015)
  1726--1733.
\newblock \href {https://doi.org/10.1109/TIM.2014.2381791}
  {\path{doi:10.1109/TIM.2014.2381791}}.

\bibitem{8059861}
G.~Jiang, P.~Xie, H.~He, J.~Yan, Wind turbine fault detection using a denoising
  autoencoder with temporal information, IEEE/ASME Transactions on Mechatronics
  23~(1) (2018) 89--100.

\bibitem{article}
J.~Liu, J.~Wang, W.~Yu, Z.~Wang, G.~Zhong, F.~He, Semi-supervised deep learning
  recognition method for the new classes of faults in wind turbine system,
  Applied Intelligence 52 (2022) 1--13.

\bibitem{https://doi.org/10.1002/acs.3685}
J.~Wang, J.~Liu, P.~Tong, W.~Yu, Z.~Wang, A new class of fault recognition
  method for wind turbine systems based on deep learning, International Journal
  of Adaptive Control and Signal Processing 37~(12) (2023) 3328--3342.

\bibitem{9327166}
Z.~Zheng, Q.~He, G.~Jiang, F.~Yin, X.~Wu, P.~Xie, Spatio-temporal
  attention-based neural network for wind turbine blade cracking fault
  detection, in: 2020 Chinese Automation Congress (CAC), 2020, pp. 7439--7444.

\bibitem{lee_transformation_2015}
J.-K. Lee, J.-Y. Park, K.-Y. Oh, S.-H. Ju, J.-S. Lee, Transformation algorithm
  of wind turbine blade moment signals for blade condition monitoring,
  Renewable Energy 79 (2015) 209--218.
\newblock \href {https://doi.org/10.1016/j.renene.2014.11.030}
  {\path{doi:10.1016/j.renene.2014.11.030}}.

\bibitem{book}
M.~Blanke, M.~Kinnaert, J.~Lunze, M.~Staroswiecki, Diagnosis and fault-tolerant
  control, 2015.
\newblock \href {https://doi.org/10.1007/978-3-662-47943-8}
  {\path{doi:10.1007/978-3-662-47943-8}}.

\end{thebibliography}

\end{document}